\documentclass{elsart3p}
\usepackage{natbib}
\usepackage{graphicx}

\newcommand{\la}{\left<}
\newcommand{\ra}{\right>}
\newcommand{\Nav}{\mbox{$\la N \ra$}}
\newcommand{\Rend}{\mbox{$R_e$}}

\newcommand{\bstar}{\mbox{$b_e$}}
\newcommand{\rhostar}{\mbox{$\rho^*$}}

\begin{document}
\runauthor{J.P.~Wittmer}
\begin{frontmatter}
\title{Are polymer melts ``ideal"?}
\author[ICS]{\underline{J.P.~Wittmer}\thanksref{X}},
\author[ICS]{P.~Beckrich},
\author[ICS]{F.~Crevel},
\author[Metz]{C.C.~Huang},
\author[ICS]{A.~Cavallo},
\author[ICS]{T.~Kreer},
\author[ICS]{H.~Meyer}
\address[ICS]{Institut Charles Sadron, 67083 Strasbourg Cedex, France}
\address[Metz]{LPMD, Inst. Physique-Electronique, Univ. Paul Verlaine-Metz,
1bd Arago, 57078 Metz cedex 3, France}
\thanks[X]{e-mail: jwittmer@ics.u-strasbg.fr}

\begin{abstract}
It is commonly accepted that in concentrated solutions or melts high-molecular 
weight polymers display random-walk conformational properties without long-range 
correlations between subsequent bonds. This absence of memory means, for instance, 
that the bond-bond correlation function, $P(s)$, of two bonds separated by $s$ 
monomers along the chain should exponentially decay with $s$. 
Presenting numerical results and theoretical arguments for both monodisperse
chains and self-assembled (essentially Flory size-distributed) equilibrium polymers 
we demonstrate that some long-range correlations remain due to self-interactions
of the chains caused by the chain connectivity and the incompressibility of the melt.
Suggesting a profound analogy with the well-known long-range velocity correlations 
in liquids we find, for instance, $P(s)$ to decay algebraically as $s^{-3/2}$.
Our study suggests a precise method for obtaining the statistical segment length 
\bstar \ in a computer experiment.
\end{abstract}
\end{frontmatter}


\section{Introduction}
\label{sec_intr}

Following Flory's ``ideality hypothesis" \cite{FloryBook},
one expects a macromolecule of size $N$ (numbers of monomers per chain) 
in a melt (disordered polymeric dense phase) to follow Gaussian
statistics \cite{DegennesBook,DoiEdwardsBook}.
The official justification of this mean-field result is that density
fluctuations are small, hence, negligible. 
One immediate consequence is of course that the mean-squared size
of a chain segment of curvilinear length $s=m-n$ scales as
\begin{equation}
R^2(s) = \la ({\bf r}_{m=n+s} - {\bf r}_n )^2 \ra_n = \bstar^2 s,
\label{eq_bstardef}
\end{equation}
at least if the two monomers $n$ and $m$ are sufficiently 
separated along the chain backbone and local correlations
may be neglected ($s\gg 1$).
Here, \bstar \ denotes the statistical segment length \cite{DoiEdwardsBook}
and ${\bf r}_i$ the position vector of monomer $i$.
Obviously, the chain end-to-end distance is then $\Rend \equiv R(s=N) = \bstar \sqrt{N}$.

Recently, this cornerstone of polymer physics has been challenged both 
theoretically \cite{SMK00,SJ03,WMBJOMMS04} and numerically 
\cite{WMBJOMMS04,CMWJB05} for three-dimensional melts and ultrathin
films \cite{SJ03,CMWJB05}.
The physical idea behind the predicted long-range intrachain correlations is
related to the ``segmental correlation hole" \cite{DegennesBook},
$\rhostar(s) \approx s/R(s)^d$, of a typical chain segment of
length $s$ in $d$ dimensions. Due to the overall incompressibility
of the melt this is thought to set an entropic penalty 
$\delta U(s) \approx \rhostar(s)/\rho$  ($\rho$ being the total density)
against bringing two adjacent chain segments together \cite{SJ03}.
The swelling of the chains caused by this repulsion is most pronounced for 
small curvilinear distances where $\delta U(s)$ is important but vanishes 
for large $s$ (see Ref.~\cite{SJ03} for logarithmic corrections relevant 
for ultrathin films, $d \rightarrow 2^+$)
at variance to the behaviour of dilute polymer chains 
in good solvent \cite{DegennesBook}.

In $d=3$ dimensions, the segmental correlation hole effect is weak,
$\delta U(s) \propto 1/\sqrt{s}$, and a standard one-loop
perturbation calculation \cite{WMBJOMMS04} can be readily 
performed following closely Edwards \cite{DoiEdwardsBook}. 
For instance, for the chain segment size this yields 
to leading order
\begin{equation}
1-  \frac{R^2(s)}{\bstar^2 s} = \frac{\kappa_R}{\sqrt{s}}
\mbox{ with } \kappa_R = \frac{\sqrt{24/\pi^3}}{\rho \bstar^3}
\label{eq_Rs}
\end{equation}
if intermediate distances larger than the monomer ($1 \ll s$) and 
smaller than the end-to-end distance ($s \ll N$) are probed.
The latter restriction avoids additional universal physics due to 
chain end effects and polydispersity 
which may be calculated in principle \cite{WMBJOMMS04}.
In the present study we do not focus on these effects.
Note that the l.h.s. should be rigorously zero for an ideal chain with statistical 
segment length \bstar \ while the r.h.s. is proportional to the perturbation
potential $\delta U(s)$.

In this paper, Eq.~(\ref{eq_Rs}) is checked numerically by
means of a well-studied lattice Monte Carlo approach \cite{BWM04}
and used to extract a precise value for the statistical 
segment size $\bstar$ for asymptotically long chains
(cf. Figs.~\ref{fig1} and \ref{fig2}).
We compare melts containing only monodisperse chains 
with systems of equilibrium polymers (EP) \cite{CC90,WMC98}
where the self-assembled chains (no closed loops being allowed)
have an annealed size-distribution of (essentially) Flory type. 
Finally, a striking demonstration of the intrachain correlations is
obtained by the power-law decay of the bond-bond correlation function
(cf. Fig.~\ref{fig3}).
We begin our discussion by briefly describing the numerical algorithms 
which allow to sample dense melts containing the large chain sizes
needed ($N \ge 1000$) for a clear-cut test of the theory.

\section{Computational questions}
\label{sec_simu}

For both monodisperse and equilibrium polymer systems we compare data obtained 
with the three dimensional bond-fluctuation model (BFM) \cite{BWM04} where each 
monomer occupies the eight sites of a unit cell of a simple cubic lattice. 
In all cases presented here cubic periodic simulation boxes of linear size $L=256$ 
containing $2^{20}$ monomers have been used. These large systems are needed to 
suppress finite size effects, especially for EP \cite{WMC98}. 
The monomer number relates to a number density $\rho=0.5/8$ where half of the 
lattice sites are occupied (volume fraction $0.5$). It has been shown that 
for this density the BFM has a low compressibility which is
comparable to real experimental melts \cite{BWM04}.
The chain monomers are connected by (at most two saturated) bonds.
In monodisperse systems adjacent monomers are permanently linked 
together (at least if only local moves through phase space are considered).
EP, on the other hand, have a finite and constant 
scission energy $E$ attributed to each bond (independent of density, 
chain length and position of the bond) which has to be paid whenever
the bond between two monomers is broken. 
(Apart from this finite scission energy for EP all our systems are perfectly 
athermal. We set $k_BT=1$ and all length scales will be given below in 
units of the lattice constant.) 
Standard Metropolis Monte Carlo is used to break and recombine the chains. 
Branching and formation of closed rings are explicitly forbidden, such that
only linear chains are present. 

The monodisperse systems have been equilibrated using a mix of local, 
slithering snake and double pivot moves \cite{BWM04}.
This allows to generate ensembles containing about $10^3$ independent 
configurations with chain length up to $N=4096$ monomers. 
We have sampled EP systems with scission energies up to $E=15$,
the largest energy corresponding to a mean chain length $\Nav \approx 6011$.
For EP only local hopping moves are needed in order to sample independent 
configurations since the breaking and recombination of chains reduces the 
relaxation times dramatically, at least for large scission-recombination 
attempt frequencies. 
In fact, all EP systems presented here have been sampled within four months 
while the $N=4096$ sample of monodisperse chains alone required about three 
years on the same XEON processor. EP are therefore very interesting from the 
computational point of view and allow for an efficient test of predictions
applying to both system classes.
%

\section{Numerical results}
\label{sec_resu}

\begin{figure}[tb]
\includegraphics*[width=0.45\textwidth]{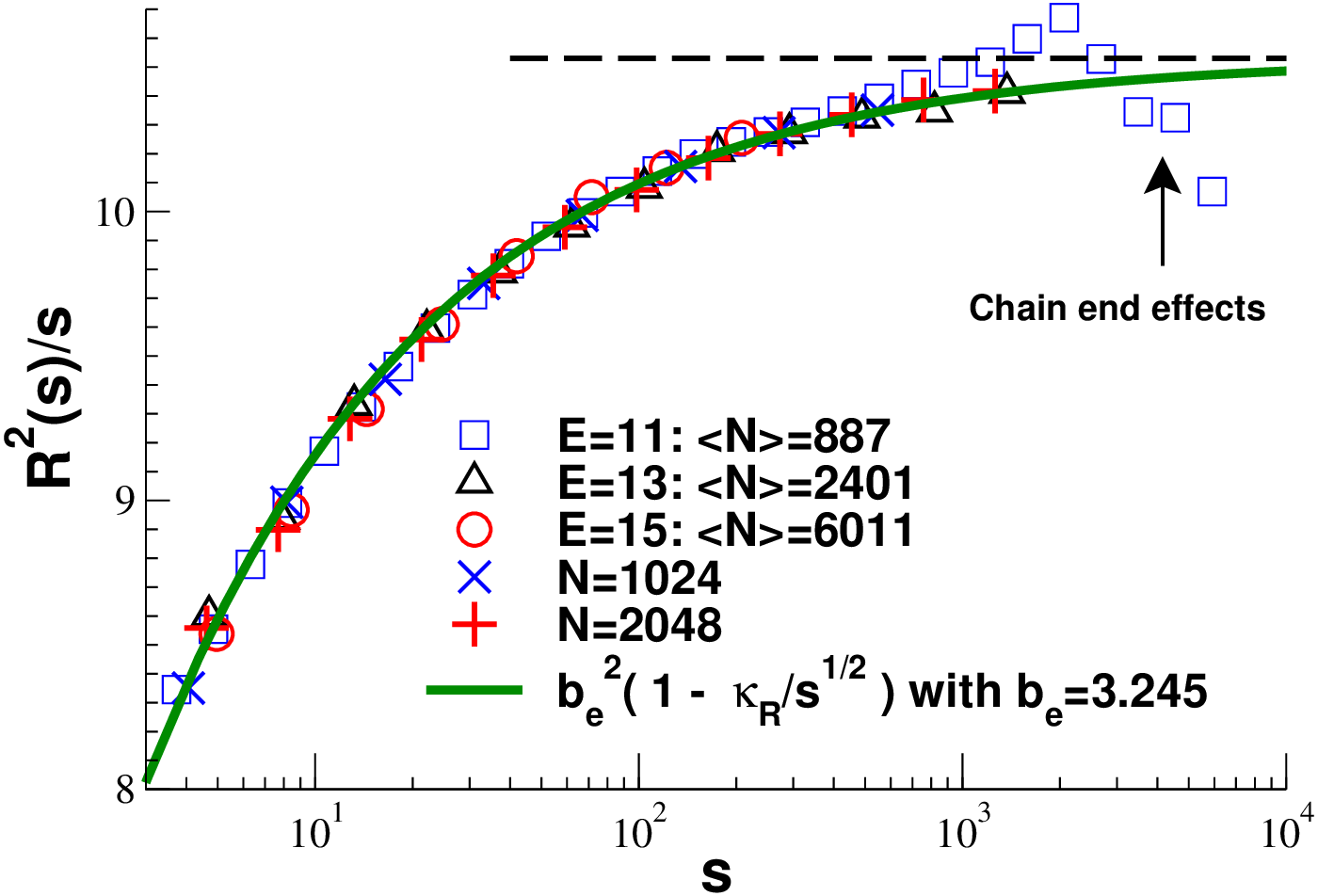}
\caption{
\label{fig1}
Curvilinear distance $R^2(s)/s$ for monodisperse chains and equilibrium
polymers (EP) obtained by Monte Carlo simulation of the bond-fluctuation
model (BFM) {\em vs.} curvilinear distance $s$.
We use log-linear coordinates to emphasize the
power-law swelling over several orders of magnitude of $s$.
The data points approach the asymptotic behaviour from below,
i.e. the chains are swollen, as predicted by Eq.~(\ref{eq_Rs}).
Provided $s \ll \Nav$, both system classes are identical.
For larger $s$ intricate chain end effects become relevant
--- especially for the polydisperse EP as indicated
for the scission energy $E=11$ ---
which are out of the scope of this study.
The averages are taken for all possible monomer pairs $(n,m=n+s)$.
The statistics deteriorates, hence, for large $s$.
}
\end{figure}

\begin{figure}[tb]
\includegraphics*[width=0.45\textwidth]{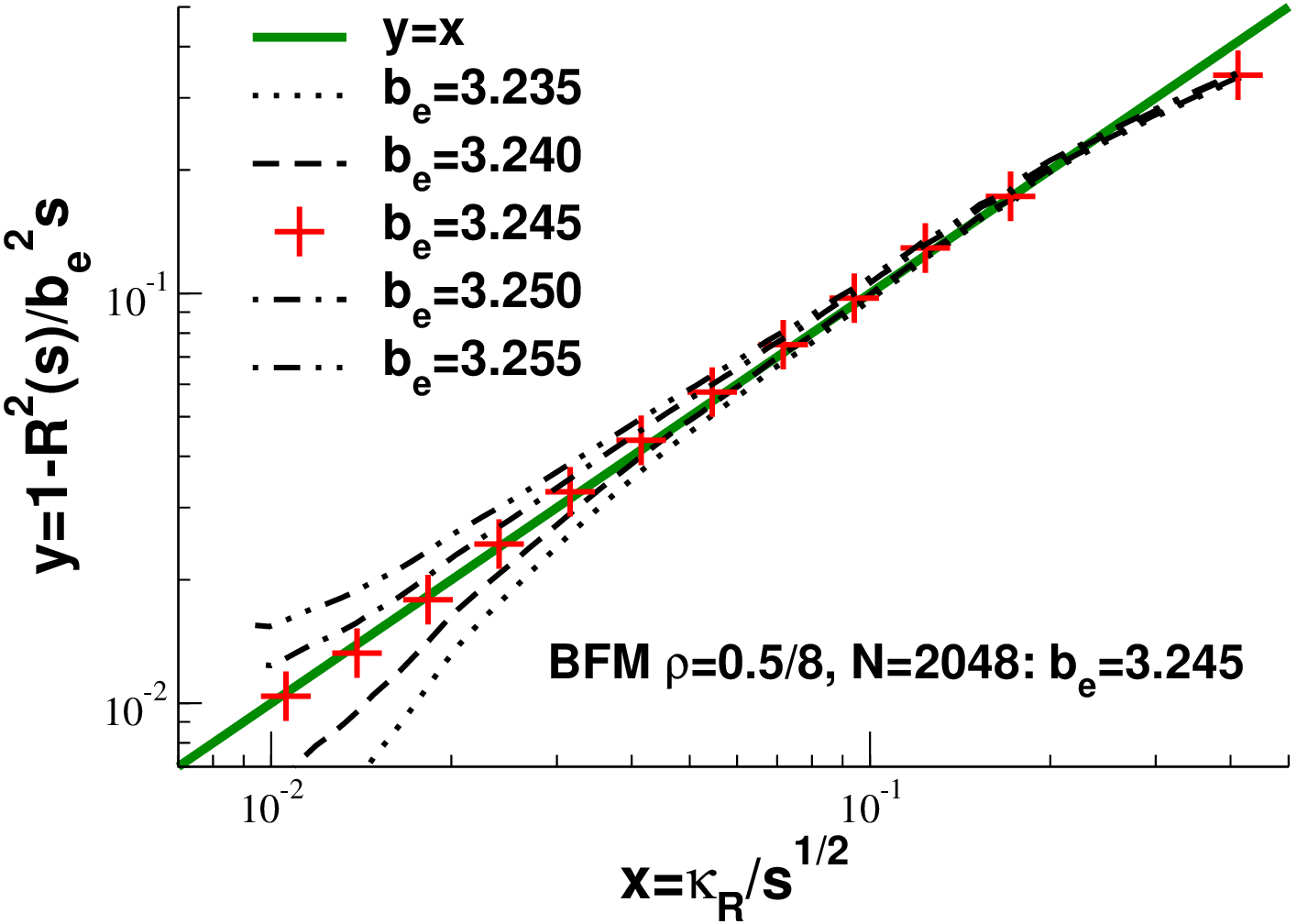}
\caption{
\label{fig2}
Replot of the segment size $R(s)$, as suggested by
Eq.~(\ref{eq_Rs}), for monodisperse chains of length $N=2048$
for different trial statistical segment lengths, as indicated.
The bold line for the value $\bstar \approx 3.245$
shows that the correction is indeed proportional to the correlation
hole, $1/\sqrt{s}$.
This procedure is very sensitive to the value chosen and allows for
a precise determination.
The same $\bstar$ allows to collapse data points from all
systems (not shown) if $s \ll \Nav$ and $100 \ll \Nav$.
}
\end{figure}

We now demonstrate numerically the long-range intramolecular correlations 
predicted by theory. Data for both monodisperse and equilibrium polymers 
(open symbols) are presented. As one expects, the size-distribution
of polymers is found irrelevant for small segments, $s \ll \Nav$.
Note that averages are taken over all possible pairs of monomers $(n,m=n+s)$ 
and, hence, the statistics deteriorates for large $s$. 
For clarity and to focus the discussion on internal correlations without
chain end effects, data points for large $s$ have been omitted for most 
of the indicated samples. For our flexible BFM chains already small curvilinear 
distances of about $s \approx 10$ are well fitted by the theory,
provided the {\em total} chain length is large, $\Nav \gg 10^3$. 

The size $R(s)$ of curvilinear chain segments of length $s$, as defined in 
Eq.~(\ref{eq_bstardef}), is presented in Figs.~\ref{fig1} and \ref{fig2}. 
The first figure clearly shows that the chains are swollen. As predicted,
the asymptotic Gaussian behaviour $R^2(s)/s \rightarrow \bstar^2$ (dashed line)
is approached from below and the deviation decays as 
$\delta U(s) \propto 1/\sqrt{s}$ (bold line). 
(The more intricate non-monotonous behaviour for large $s$ for both system classes, 
especially for EP, is indicated for one example with scission energy $E=11$.)

Computationally, $R(s)$ is an important quantity since it allows for 
the precise 
determination of the statistical segment length \bstar \ by extrapolation
to asymptotically large segment sizes by means of the one parameter fit 
suggested by Eq.~(\ref{eq_Rs}). The bold line indicated corresponds
to $\bstar = 3.245$ which nicely fits the data over several decades in $s$. 
Note that a systematic {\em underestimation} of the true statistical segment length 
would be obtained by taking simply the largest $R^2(s)/s \approx 3.23^2$ value 
available, say, for monodisperse chains of length $N=2048$.
A better representation for performing the fit is to plot $R^2(s)/s$ {\em vs.} 
$1/\sqrt{s}$ in linear coordinates (not shown). This has the
disadvantage that data points with $s \gg 10$ are less visible. 
A much more precise representation is given in Fig.~\ref{fig2} where we plot 
in logarithmic coordinates for different trial values of \bstar \
the l.h.s.  {\em vs.} the r.h.s. of Eq.~(\ref{eq_Rs}), 
i.e. the non-Gaussian deviations against the depth of the segmental correlation hole.
Since both sides of the relation must be identical (neglecting higher order 
perturbation corrections) the fit value should yield a data collapse on the
bisection line (bold line). Note that our best value yields a collapse over 
more than an order of magnitude {\em without} curvature.

\begin{figure}[tb]
\includegraphics*[width=0.45\textwidth]{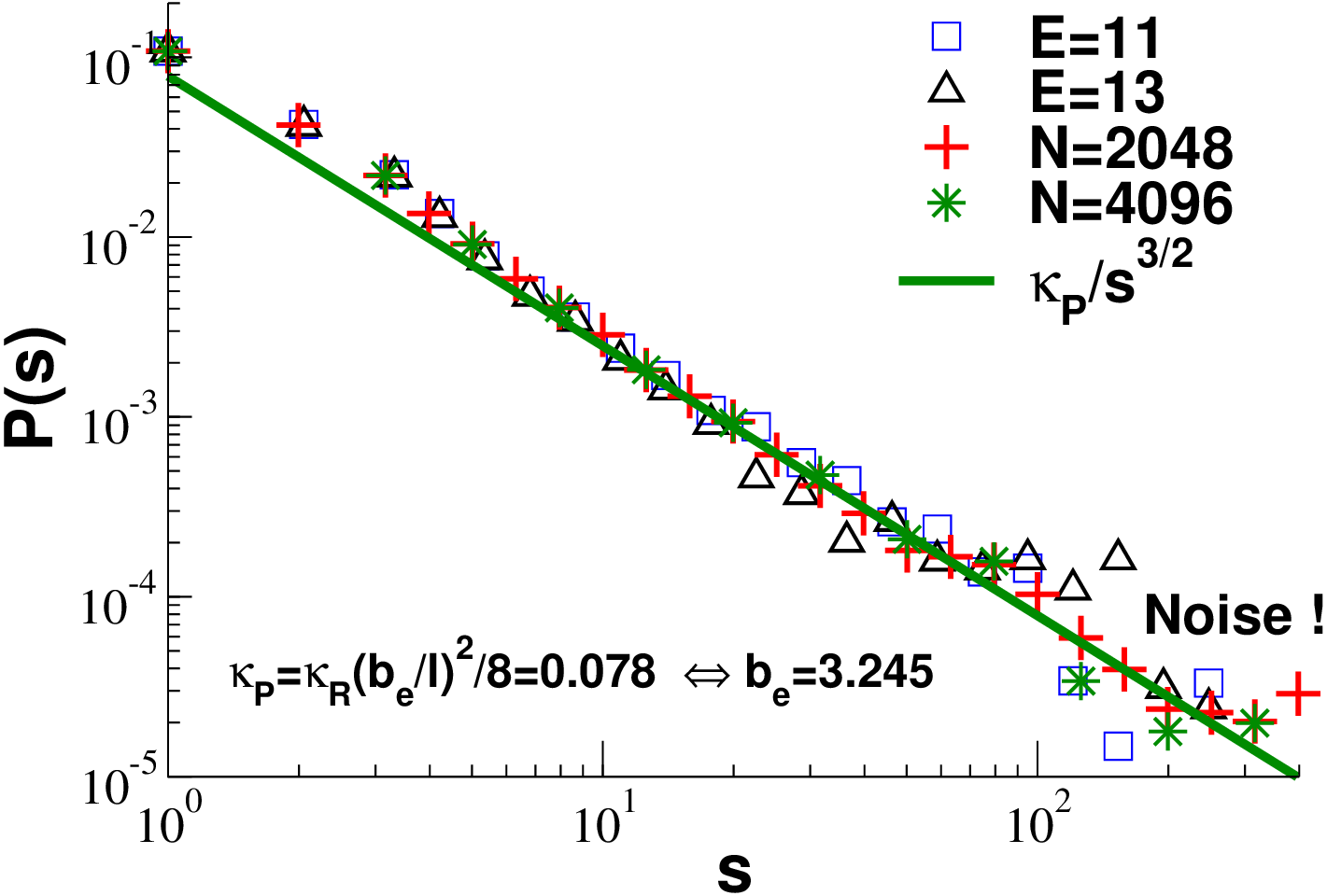}
\caption{The bond-bond correlation function $P(s)$ for both system
classes considered (at number density $\rho=0.5/8$) as a function
of the curvilinear distance $s$. Provided that $1 \ll s \ll \Nav$,
all data sets collapse on the power-law slope (bold line) predicted
by Eq.~(\ref{eq_Ps}).
The coefficient $\kappa_P$ indicated is consistent with a statistical
segment length $\bstar \approx 3.245 \pm 0.005$.
\label{fig3}
}
\end{figure}

An even more striking violation of Flory's ideality hypothesis may 
be obtained by computing the bond-bond correlation function,
$P(s)= \la {\bf l}_{m=n+s} \cdot {\bf l}_n \ra_n / l^2$.
Here, ${\bf l}_i = {\bf r}_{i+1}- {\bf r}_i$ denotes the
bond vector between two adjacent monomers $i$ and $i+1$
and $l(\rho=0.5/8) \approx 2.634$, the (mean-squared) bond length. 
It can be shown that $P(s)$ must decay exponentially if no long-range
memory effects were present due to the multiplicative loss of information 
transferred recursively along the chain \cite{FloryBook}. 
The bond-bond correlation function is in fact the second derivative of $R^2(s)$ 
with respect to $s$
\begin{equation}
P(s) \ l^2 = - \frac{1}{2} \frac{d^2}{ds^2} R^2(s),
\label{eq_RsPsconnect}
\end{equation}
hence, it allows us to probe the non-Gaussian corrections directly without 
the trivial ideal contribution \cite{WMBJOMMS04}.
(We remember that the velocity correlation function of a fluid is similarly 
related to the second derivative with respect to time of the mean-square 
displacement.)
This relation together with Eq.~(\ref{eq_Rs}) immediately yields
an algebraical decay 
\begin{equation}
P(s) = \frac{\kappa_P}{s^{3/2}}
\mbox{ with }
\kappa_P = \kappa_R (\bstar/l)^2 /8 
\label{eq_Ps}
\end{equation}
of the bond-bond correlation function for dense solutions and melts,
rather than the exponential cut-off expected from Flory's 
hypothesis.\footnote{
Interestingly, the scaling of the bond-bond correlation function is related to 
the long-time power-law correlations of the velocity-correlation function 
found in dense fluids nearly fourty years ago \cite{Alder}. 
Albeit the {\em constraints} which cause the self-interactions
are different (momentum conservation for fluid particles,
incompressibility for polymer chains) the weight with which
these constraints increase the stiffness of the random walker
is in both cases proportional to the return probability,
$1/s^{d/2}$, of a random walker in $d$ dimensions.
}
This prediction is perfectly confirmed by the data collapse presented
in Fig.~\ref{fig3} for both system classes. 
In principle, \bstar \ may also be measured from the power-law amplitude, 
however, to less accuracy than the previous method (Fig.~\ref{fig2}).

\section{Conclusion}
\label{sec_conc}

In this study we have challenged the famous ideality hypothesis suggested 
half a century ago by Flory. Testing recent theoretical predictions 
\cite{SJ03,WMBJOMMS04} by means of extensive lattice Monte Carlo 
simulations \cite{BWM04} of linear polymer melts having either a quenched 
and monodisperse or an annealed size-distribution we provide strong evidence 
that long-range correlations exist in dense polymer melts which tend to
swell the chains.
Their most striking effect is the power-law asymptote for the bond-bond
correlation function suggesting a profound analogy with the well-known 
long-range velocity correlations in liquids \cite{Alder}.
We finally emphasize that our results rely on {\em generic} physics which 
should apply to all polymer melts containing {\em long} and preferentially 
{\em flexible} chains.

\ack
We thank A.~Johner (ICS, Strasbourg, France), 
S.P.~Obukhov (Gainesville, Florida), N.-K.~Lee (Seoul, Korea),
H.~Xu (Metz, France) and J.-P.~Ryckaert (Bruxelles, Belgium) 
for discussions, and J. Baschnagel for critical reading of the manuscript.
Computer time by the IDRIS (Orsay) is also acknowledged.
We are indebted to the DFG (KR 2854/1-2), the Universit\'e Louis Pasteur, 
and the ESF STIPOMAT programme for financial support.
 

\end{document}